\documentclass[useAMS,usenatbib]{mn2e}
\usepackage{lscape}
\usepackage{rotating}
\usepackage{longtable}

\usepackage[T1]{fontenc}
\usepackage[utf8x]{inputenc}
\usepackage{graphicx}
\usepackage{amsmath}
\usepackage{natbib}

\title[ICM emission in high z FRIs]{The X--ray properties of high-z
  FRI candidates in  the COSMOS field}  

\author[E. Tundo, P. Tozzi and M. Chiaberge]{E. Tundo$^{1}$\thanks{mail: tundo@oats.inaf.it (ET);
    tozzi@oats.inaf.it (PT), marcoc@stsci.edu (MC)},
  P. Tozzi$^{1,}$$^{2}$ and M. Chiaberge$^{3,}$$^{4}$ \\
$^{1}$INAF - Osservatorio Astronomico di Trieste, Via Tiepolo 11,
I-34143 Trieste, Italy\\
$^{2}$INFN -- Istituto Nazionale di Fisica Nucleare, Sezione di Trieste, Italy\\
$^{3}$Space Telescope Science Institute, 3700 San Martin
  Drive, Baltimore, MD 21218\\
$^{4}$INAF - IRA, Via P. Gobetti 101, Bologna I-40129, Italy}
\begin{document}

\def\ltsima{$\; \buildrel < \over \sim \;$}
\def\simlt{\lower.5ex\hbox{\ltsima}}
\def\gtsima{$\; \buildrel > \over \sim \;$}
\def\simgt{\lower.5ex\hbox{\gtsima}}

\date{October 6th, 2011}

\pagerange{\pageref{firstpage}--\pageref{lastpage}} \pubyear{0000}

\maketitle

\label{firstpage}

\begin{abstract}
We report the X--ray analysis of a sample of candidate high-redshift
(1$<$z$<$2) FRI sources from the sample of Chiaberge et al. (2009),
observed in the Chandra COSMOS field (C-COSMOS). Our main goals are to
study their nuclear properties by means of unresolved X--ray emission,
and to constrain the presence of clusters surrounding the FRI sources
from the diffuse X--ray emission by the associated hot plasma. Among
19 FRI candidates, 6 have an X--ray unresolved counterpart in the
C-COSMOS catalog. Two additional sources are not present in the
C-COSMOS catalog but are clearly detected in the Chandra
images. X--ray spectral analysis, when possible, or hardness ratio of
the stacked emission from X--ray detected sources, suggest that some
of them have significant intrinsic absorption ($N_H \simgt 10^{22}$
cm$^{-2}$), and high X--ray luminosities with respect to local FRIs.
From the stacking analysis of the 11 non-detected sources, however, we
find evidence for unresolved soft X--ray emission and no detected hard
emission, suggesting an unabsorbed spectrum. Therefore, the X--ray
properties vary significantly from source to source among these FRI
candidates. From the analysis of the stacked images of all 19 FRI
candidates we can rule out the presence of virialized haloes with
temperatures larger than 2-3 keV; however, the upper limit on the
average extended emission is still consistent with the presence of
$\sim$1$-$2 keV hot gas.
\end{abstract}


\section{Introduction}

Among the most energetic phenomena in the universe, radio galaxies are
expected to play a major role in the study of several topics in
astrophysics, such as accretion onto supermassive black holes (SMBH),
the associated formation of relativistic jets (e.g. Blandford \& Payne
1982; Livio 1999), the feedback processes of an accreting SMBH in the
star formation history of a galaxy (e.g. Hopkins et al. 2006) and the
role of the Active Galactic Nuclei (AGN) in injecting energy in the
intracluster medium (Fabian, Celotti \& Erlund 2006). Only recently the
possibility of using the connection between Fanaroff-Riley I radio
galaxies (FRIs; Fanaroff \& Riley~1974) with elliptical massive
galaxies in cluster (Owen 1996; Zirbel 1996) has been explored in the
frame of the search of clusters in the redshift range 1$<$z$<$2 (see
Chiaberge et al. 2009, Chiaberge et al. 2010).

Following the AGN unification scheme (Antonucci~1993; Urry \&
Padovani~1995 and reference therein), radio galaxies are the
mis-oriented parent population of jet-dominated blazars, and thus
correspond to large viewing angles to the jet/dusty torus axes.
Emission from the misaligned sources should be significantly obscured
by dust and gas associated with the torus.  The centrally-brightened
FRI galaxies should be the misaligned versions of BL Lac objects,
while most edge-brightened FRIIs are unified with powerful
quasars. FRI galaxies typically have a radio power lower than that of
FRII sources, with the FRI$/$FRII break set at $L_{178MHz}\sim 2
\times 10^{33}$ erg s$^{-1}$ Hz$^{-1}$ or $L_{1.4GHz} \sim 4 \times 10
^{32}$ erg s$^{-1}$ Hz$^{-1}$.  A more accurate scheme is given by
Jackson \& Wall (1999), who claim that there are two parent
populations, the high radio--power FRII radio galaxies which are the
parents of all radio quasars and some BL Lac-type objects, and the
moderate--radio--power FRI radio galaxies which are the parents of the
majority of BL Lac--type objects. The transition between FRIs and FRIIs
is anyway rather smooth and both radio morphologies are present in the
population of sources around the break. The FRI$/$FRII break (at low
redshift) also depends on the luminosity of the host galaxy, as shown
by Owen \& Ledlow (1994) and Ledlow \& Owen (1996). However, it is
still unclear whether or not that might simply be a result of
selection effects (Scarpa \& Urry 2001).

As a further complication, objects of intermediate radio structure do
exist (e.g. Capetti, Fanti \& Parma 1995), and a class of hybrid double
sources, with a FRI jet on one side and a FRII lobe on the other, was
also unveiled by Gopal-Krishna \& Wiita (2000). This supports
explanations for the FR dichotomy based upon different jet interaction
with the Inter Galactic Medium of the same nuclear engine (Kaiser \&
Alexander 1997, Gawro{\'n}ski et al. 2006, Evans et al. 2008).  On the
other hand, in some models the different radio power is expected to be
associated to different accretion modes, with FRIIs being powered by a
standard accretion disc and FRIs by an advection dominated accretion
flow (see Reynolds et al. 1996; Ghisellini \& Celotti 2001).

Apart from their intrinsic properties, FRIs and FRIIs differ also in
their typical environment.  The FRI and FRII host galaxies are
mostly elliptical massive galaxies in the local universe (e.g. Zirbel
1996; Donzelli et al. 2007), and thus are associated with the most
massive black holes in the local universe through the
M$_\bullet$-$\sigma_\ast$ relation (Ferrarese \& Merrit 2000, Gebhardt
et al. 2000, Tremaine et al. 2002). However, FRIs are usually located
at the centre of massive clusters (see e.g. Owen 1996, for a
review). For example, 94\% of the radio sources in the central regions
of Abell clusters are FRIs (Ledlow \& Owen~1996).  On the contrary,
FRIIs at low redshift are generally found in regions of lower
density, with an increasing fraction of FRII residing in rich groups
or clusters only at redshift higher than $\sim 0.5$ (Prestage \&
Peacock 1988; Zirbel 1996; Hill \& Lilly 1991; Best 2007). In addition
FRII host galaxies show an optical morphology often associated with
violent galaxy encounters, with a low percentage of smooth ellipticals
(Zirbel 1996).

In this framework, FRIs can be successfully used as a tool in the
search for high redshift clusters.  However, so far this aspect has
not been exploited properly (see Chiaberge et al. 2009).  The optical
search for clusters of galaxies at z$>$1 has proven to be particularly
difficult, mainly because of the reduced contrast between cluster
members and field galaxies.  In order to tackle this difficulties,
high-z clusters have been recently searched using the so-called red
sequence technique (Gladders \& Yee 2000).  With a similar approach
extended to longer wavelengths, Eisenhardt (2008) presented a sample
of clusters of galaxies from the Spitzer Infrared Array Camera Shallow
Survey. Among their 335 cluster candidates, 18 have confirmed
spectroscopic redshift z$\sim$1-1.4, and a few have photometric
redshift (obtained using photometry in 4 bands, B$_W$, R, I, and
3.6$\mu$m) in the range z$_{phot}\sim$1.5-2. A similar effort is
being pursued by the Spitzer Adaptation of the Red-Sequence Cluster
Survey (SpARCS
\footnote{http://www.faculty.ucr.edu/\~gillianw/SpARCS/}) project.
Their cluster detection algorithm lead to about two hundred high z
cluster candidates, and the follow-up and analysis phase,
which will last for several years, had spectroscopically confirmed
about a dozen clusters at z$>$1 so far (see Wilson et al. 2009; Muzzin et
al. 2009; Demarco et al. 2010).  Turning to the X--ray window explored
by the Chandra, XMM and Swift satellites, several new detections are found
based on the compilation of serendipitous medium and deep--exposure
extragalactic pointings not associated to previously known X--ray
clusters.  Among them, the XMM Deep Cluster Survey (XDCP
\footnote{http://www.xray.mpe.mpg.de/theorie/cluster/XDCP/xdcp\_index.html},
Fassbender et al. in prep. that has so far spectroscopically confirmed
18 clusters at z$>$0.8), the ChaMP Serendipitous Galaxy Cluster Survey
(Barkhouse et al. 2006), and the Swift-XRT Cluster Survey (SXCS, Tundo
et al., Moretti et al. in prep.).  In addition, there are also ongoing
dedicated surveys like the XMM-BCS cluster survey (Suhada et al. 2011,
in prep.), and the XMM Large Scale Structure Survey (XMM-LSS; Pierre,
Valtchanov \& Refregier~2002).  So far, the highest confirmed redshift
for an X--ray detected cluster is z$\sim$1.62 (Tanaka et al. 2010),
with a few extreme candidates at z$\sim$2 (see, e.g., Andreon \&
Huertas-Company 2011).  To summarize, the present-day limit for the
detection of well characterized, virialized cluster of galaxies, is
z$\simeq$1.5.
 
Given the properties of the environment of local FRIs and their
association with massive ellipticals in clusters at z$\sim$0, FRIs may
help in the search of high redshift clusters at z$\geq$1.5. In
Chiaberge et al. (2009; 2010) the selected sample of FRIs in the
COSMOS field (Scoville et al. 2007) is used to search for cluster
candidates in this relatively unexplored redshift range.  With this
study we want to exploit the Chandra X--ray data in the COSMOS field
to investigate the X--ray properties of the FRI sample of Chiaberge et
al. (2009).  Thanks to the deep C-COSMOS X--ray images, this work
carries out a study to investigate both the intrinsic properties of
FRIs at high redshift and the presence of Intra Cluster Medium
emission associated to possible clusters hosting the FRIs.

The Paper is organized as follows.  In \S 2 we present the sample; in
\S 3 we study the photometric X--ray properties of the FRI candidates
and the spectral analysis of the 8 sources with unresolved X--ray
counterparts.  In \S 4 we explore the possibility of detecting
extended emission with a stacking analysis.  Finally in \S 5 we
summarize our results. We use a standard flat $\Lambda$CDM
cosmological model with the best-fit parameters from the seven years
WMAP data $\Omega_m=$0.272, $\Omega_\Lambda=$0.728, and $H_0=$70.4 km
s$^{-1}$ Mpc$^{-1}$ (Komatsu et al. 2011).

\begin{table}
 \centering
 \begin{minipage}{85mm}
\caption{FRI sources in the C-COSMOS field: (1) FRI candidates Id
  (Chiaberge et al. 2009); (2) C-COSMOS X--ray catalog Id (if present)
  (3),(4) RA DEC coordinates of the FRI candidates; (5) photometric
  redshift (Ilbert et al. 2009).}
\label{FRI_src}
\begin{tabular}{@{}ccccc@{}}
\hline
 \bf FRI & \bf C-COSMOS \footnote{FRI sources 1 and 4, while not present in the C-COSMOS catalog, have a robust X--ray counterpart.}& \bf Ra & \bf Dec   &\bf z$_{phot}$ \footnote{$\dagger$: spectroscopic redshift  from the zCOSMOS catalog (Lilly et al. 2007); $\ddagger$:  spectroscopic redshift from  Magellan-COSMOS (Trump et al. 2007); $\ast$ photometric redshift from Salvato et al. (2011).}  \\   \hline
\bf 01  &      & 150.20744  & 2.28187  & 0.883$^{+0.004}_{-0.004}$ $\dagger$ $\ddagger$ \\ 
\bf 03  & 451  & 150.00253  & 2.25863  & 2.09$^{+0.01}_{-0.01}$  $\ast$ \\ 
\bf 04  &      & 149.99153  & 2.30277  & 1.45$^{+0.04}_{-0.02}$  \\ 
\bf 05  & 1091 & 150.10612  & 2.01447  & 1.88$^{+0.03}_{-0.05}$  $\ast$  \\ 
\bf 07  &      & 150.26883  & 2.03435  & 0.94$^{+0.02}_{-0.02}$  \\ 
\bf 11  &      & 150.07816  & 1.89855  & 1.31$^{+0.13}_{-0.08}$  \\ 
\bf 13  & 1242 & 149.97784  & 2.50420  & 1.171$^{+0.04}_{-0.06}$  $\ast$ \\ 
\bf 16  &      & 150.53772  & 2.26735  & 0.969$^{+0.001}_{-0.001}$ $\ddagger$  \\ 
\bf 18  & 427  & 149.69325  & 2.26746  & 0.904$^{+0.04}_{-0.06}$  $\ast$  \\ 
\bf 20  &      & 149.83209  & 2.56954  & 0.99$^{+0.02}_{-0.02}$  \\ 
\bf 22  &      & 149.89508  & 2.62921  & 1.79$^{+0.03}_{-0.02}$  \\ 
\bf 26  &      & 149.62114  & 2.09198  & 1.19$^{+0.01}_{-0.01}$  \\ 
\bf 28  &      & 149.60064  & 2.09186  & 1.23$^{+0.11}_{-0.06}$  \\ 
\bf 29  &      & 149.64587  & 1.95297  & 1.59$^{+0.18}_{-0.15}$  \\ 
\bf 30  &      & 149.61542  & 1.99105  & 0.90$^{+0.02}_{-0.01}$  \\ 
\bf 31  &      & 149.61916  & 1.91636  & 0.913$^{+0.001}_{-0.001}$ $\dagger$ $\ddagger$ \\ 
\bf 32  &      & 149.66830  & 1.83797  & 0.58$^{+0.01}_{-0.22}$  \\ 
\bf 52  & 701  & 149.90590  & 2.39647  & 0.74$^{+0.01}_{-0.01}$  $\ast$  \\
\bf 66  & 669  & 149.95946  & 1.80149  & 0.68$^{+0.01}_{-0.01}$  $\ast$  \\  \hline
\end{tabular}
\end{minipage}
\end{table}

\section[]{The sample}

Chiaberge et al. (2009) selected a sample of 37 high redshift
(1$<$z$<$2) FRI candidates in the COSMOS field.  They proceeded in four
steps, under few basic assumptions. The two main assumptions are: the
FRI$/$FRII break in radio power per unit frequency does not change with
redshift; the photometric properties of the hosts of FRIs at 1$<$z$<$2
are similar to those of FRIIs within the same redshift bin, as in the
case of local radio galaxies (e.g. Donzelli et al. 2007).

They selected FIRST radio sources (see Becker, White \& Helfand 1995) in the
COSMOS field according to their 1.4 GHz flux.  Sources are considered
FRI candidates when the 1.4 GHz flux falls within the range of fluxes
expected for FRIs at 1$<$z$<$2.  This criterion is $1<F_{1.4}<13$
mJy. From the analysis of the VLA-COSMOS (Schinnerer et al. 2007)
images, sources with FRII radio morphology were excluded.  Bright
(m$_I>$ 22) galaxies were rejected as likely low−z galaxies with
intrinsically faint radio emission (e.g. nearby starbursts), and
U-band dropouts were rejected as likely to be at z$>$2.5. This lead to
the final sample of 37 FRI candidates.

In this work we search for the X--ray counterparts of the FRI
candidates included within the Chandra COSMOS Survey (C-COSMOS, Elvis
et al. 2009).  The C-COSMOS is a large, 1.8 Ms, Chandra program that
has imaged the central 1 deg$^2$ of the COSMOS field (centred at
10$^h$00$^m$28.6$^s$, +02°12'21.0'', and overlapping with a region of
2 deg$^2$ imaged with XMM-Newton, see Cappelluti et al. 2009).  This
results in an average exposure of $\sim$160~ksec in the inner 0.5
deg$^2$ region, and an outer 0.5~deg$^2$ area with an effective
exposure of $\sim$ 80~ksec. The limiting fluxes for point source
detection are 1.9$\times$10$^{-16}$erg cm$^{-2}$ s$^{-1}$ in the soft
(0.5-2 keV) and 7.3$\times$10$^{-16}$ erg cm$^{-2}$ s$^{-1}$in the
hard (2-10 keV) band.
 
XMM--COSMOS (Hasinger et al. 2007) would indeed offer a bigger
effective area than Chandra.  However, identification of potential
extended sources, in particular at medium and high redshift, is more
difficult than with Chandra, due to the size of the XMM Point Spread
Function (PSF, whose Half Energy Width is 15'' at the aimpoint) and
its degradation as a function of the off--axis angle.  Moreover, the
sharp PSF of Chandra makes easier to identify faint point sources
which otherwise may be confused with the possible diffuse emission.
Therefore, we decide to limit our study to the Chandra data only.

Only 19 out of the 37 FRI candidates are included in the C-COSMOS
field.  Starting from the C-COSMOS source list (Elvis et al. 2009), we
searched for all the possible X--ray counterparts within a matching
radius of 2 arcsec.  The number of expected false matches is
negligible thanks to the low number density of the X--ray and radio
sources at the limiting fluxes. Table \ref{FRI_src} lists the FRI
candidates studied in this work, with their X--ray counterparts when
present.  We find the X--ray counterparts in the C-COSMOS point source
catalog for 6 radio sources. In addition, the two radio sources FRI 1
and 4, have highly significant X-ray counterparts which are not listed
in the C-COSMOS catalog. The photometric redshifts are given by Ilbert
et al.~(2009) for all the sources without a C-COSMOS counterpart.  For
three sources we also have a spectroscopic redshift from the zCOSMOS
catalog (Lilly et al. 2007) or the Inamori Magellan Area Camera and
Spectrograph instrument on the Magellan telescope (Trump et
al. 2007). For sources with a C-COSMOS counterpart we use the
photometric redshift of Salvato et al. (2011).

\section{X--ray Analysis}

\begin{table*}
 \centering
\begin{minipage}{180mm}
\caption{X--ray aperture photometry for FRI candidates: (1) FRI
  candidate Id (Chiaberge et al. 2009); (2),(3) soft and hard net
  counts with 1 $\sigma$ Poissonian error bars; (4),(5) soft and hard Signal to
  Noise Ratio; (6) Hardness Ratio; (7),(8) observed soft and hard flux
  corresponding to an average spectral slope of $\Gamma_{av} = 1.4$;
  (9),(10) observed soft and hard luminosities. X--ray fluxes and
  luminosities for FRI sources with no X--ray counterparts correspond
  to a 3 $\sigma$ Poissonian upper limit. Uncertainties on luminosities include
  also the error in the z$_{phot}$.  }
\label{photometry_X}
\begin{tabular}{@{}cccccccccc@{}}\hline
 \bf FRI Id & \bf Cts$_S$  & \bf Cts$_H$ & \bf  S/N$_S$ &\bf S/N$_H$  & \bf HR  &  {\bf S$_{0.5-2}$ } &  {\bf S$_{2-10}$  } &  {\bf  L$_{0.5-2}$}  &  {\bf L$_{2-10}$}  \\ \hline
 \bf        & & & & &   & \multicolumn{2}{c} {\bf 10$^{-16}$ erg cm$^{-2}$ s$^{-1}$  } & \multicolumn{2}{c|} {\bf 10$^{42}$ erg s$^{-1}$  } \\  \hline

\bf 07   & $<$ 9.9  & $<$ 12.4   & -- & -- & --  &  $<$3.1   &  $<$18.3   & $<$  1.8$\pm$ 0.1  & $<$  10.7$\pm$  0.6 \\ 
\bf 11   & $<$ 8.2  & $<$ 16.4   & -- & -- & --  &  $<$2.8   &  $<$25.6   & $<$  3.9$\pm$ 0.9  & $<$  36.4$\pm$  7.5 \\  
\bf 16   & $<$ 6.9  & $<$  8.2   & -- & -- & --  &  $<$5.4   &  $<$29.8   & $<$  3.5$\pm$ 0.02 & $<$  18.9$\pm$  0.1 \\  
\bf 20   & $<$11.9  & $<$  9.1   & -- & -- & --  &  $<$7.7   &  $<$27.5   & $<$  5.3$\pm$ 0.2  & $<$  18.8$\pm$  0.7 \\ 
\bf 22   & $<$ 8.2  & $<$  9.1   & -- & -- & --  &  $<$8.5   &  $<$34.4   & $<$ 28.7$\pm$ 1.5  & $<$ 148.7$\pm$  8.0 \\ 
\bf 26   & $<$ 7.0  & $<$  8.3   & -- & -- & --  &  $<$4.6   &  $<$25.1   & $<$  5.1$\pm$ 0.1  & $<$  28.1$\pm$  0.6 \\ 
\bf 28   & $<$ 9.2  & $<$ 11.3   & -- & -- & --  &  $<$6.3   &  $<$35.5   & $<$ 57.2$\pm$ 4.5  & $<$ 325.3$\pm$ 23.5 \\ 
\bf 29   & $<$ 4.0  & $<$ 11.9   & -- & -- & --  &  $<$1.9   &  $<$26.4   & $<$  4.5$\pm$ 1.0  & $<$  63.2$\pm$ 15.5 \\ 
\bf 30   & $<$ 8.2  & $<$  8.1   & -- & -- & --  &  $<$5.3   &  $<$24.6   & $<$  2.8$\pm$ 0.1  & $<$  12.8$\pm$  0.5 \\ 
\bf 31   & $<$ 8.2  & $<$  9.2   & -- & -- & --  &  $<$8.9   &  $<$46.3   & $<$  4.8$\pm$ 0.3  & $<$  24.9$\pm$  0.5 \\ 
\bf 32   & $<$ 7.0  & $<$ 11.3   & -- & -- & --  &  $<$5.1   &  $<$38.2   & $<$  0.8$\pm$ 0.3  & $<$   6.3$\pm$  3.5 \\\hline 
\bf 01   &    5.4$^{+  4.4}_{-  3.6}$ &   8.9  $^{+ 5.6}_{-  4.6}$  &  1.55  &  1.89 &  0.25  $^{+0.5}_{-0.7}$  &  1.6$\pm$1.0 &  12.8$\pm$ 6.6 &   0.8$\pm$ 0.5  &    6.3$\pm$  3.4 \\ 
\bf 03   &  35.73$^{+ 7.3}_{- 6.3}$   & 85.06  $^{+10.8}_{-  9.8}$  &  5.65  & 8.73  &  0.41  $^{+0.2}_{-0.2}$  & 12.4$\pm$2.1 & 137.8$\pm$15.6 &  52.8$\pm$11.2  &  583.2$\pm$102.9 \\ 
\bf 04   &    9.0$^{+  4.3}_{-  3.3}$ &   3.7  $^{+ 3.8}_{-  2.8}$  &  2.72  &  1.39 & -0.41  $^{+0.8}_{-0.5}$  &  3.1$\pm$1.0 &   6.6$\pm$ 4.0 &   5.8$\pm$ 2.0  &   12.4$\pm$  8.0 \\ 
\bf 05   &  15.27$^{+  5.4}_{-  6.1}$ & 16.22  $^{+ 6.0}_{-  5.0}$  &  3.50  & 3.31  &  0.03  $^{+0.4}_{-0.5}$  &  5.2$\pm$1.4 &  27.1$\pm$ 7.4 &  19.9$\pm$ 5.7  &  103.2$\pm$ 30.0 \\
\bf 13   &  12.43$^{+  5.0}_{-  4.0}$ & 13.94  $^{+ 5.9}_{-  4.9}$  &  3.11  & 2.85  &  0.06  $^{+0.5}_{-0.5}$  &  3.8$\pm$1.2 &  19.3$\pm$ 6.9 &   4.0$\pm$ 1.3  &   20.4$\pm$  7.5 \\
\bf 18   &  10.63$^{+  4.4}_{-  3.4}$ &  0.6   $^{+ 2.9}_{-  0.6}$   &  3.07  & 0.29  & -0.90 $^{+0.7}_{-0.1}$  &  7.8$\pm$2.5 &   0.7$\pm$ 7.3 &   4.0$\pm$ 1.3  &    0.4$\pm$  3.5 \\
\bf 52   &  29.80$^{+ 6.7}_{-  5.7}$  & 14.65  $^{+ 5.7}_{-  4.7}$  &  5.19  & 3.12  & -0.34  $^{+0.4}_{-0.3}$  & 10.6$\pm$2.0 &  23.7$\pm$ 7.9 &   3.3$\pm$ 0.6  &    7.4$\pm$  2.3 \\ 
\bf 66   &   1.13$^{+  2.7}_{-  0.9}$ &  9.14  $^{+ 4.6}_{-  3.6}$  &  0.65  & 2.54  &  0.78  $^{+0.2}_{-0.9}$  &  0.7$\pm$1.1 &  27.3$\pm$10.7 &   0.2$\pm$ 0.3  &    6.8$\pm$  2.7 \\ \hline
\end{tabular}
\end{minipage}
\end{table*}

The X--ray data are processed in a standard way using the {\tt ciao
  4.3} software and the calibration database CALDBv4.4.5.  We obtained
the merged image of the entire C-COSMOS field, and performed aperture
photometry on the radio position for our sample of FRI candidates.
The results are given in Table \ref{photometry_X} for the soft (0.5-2
keV) and hard (2-10 keV) X--ray bands. The net counts and the
signal-to-noise ratio (S/N) are computed within an extraction radius
of 5 arcsec unless this radius overlap with a nearby bright source.
We use a smaller aperture only for FRI 4 whose extraction radius was
set to 3.5 arcsec. The quoted uncertainties represent the 84.13\%
upper and lower confidence level assuming a Poisson distribution, that
correspond to 1 $\sigma$ level in the Gaussian statistics (see Gehrels
1986).  For sources without an X--ray counterpart we report the
corresponding 3 $\sigma$ upper limit (in these cases S/N and hardness
ratio are not defined). Count rates and energy fluxes in the soft and
hard bands are corrected for vignetting using the composite
monochromatic exposure maps computed at 1.5 and 4.5 keV,
respectively. The hardness ratio is computed as $HR=(H-S)/(H+S)$,
where H and S are the aperture photometry in the hard and soft band,
respectively. Approximate energy fluxes are computed directly from the
measured count rate as in Rosati et al. (2002), using conversion
factors corresponding to a slope of $\Gamma_{av}$ = 1.4, that
represents the average spectral shape for X--ray sources in the
$10^{-16}-10^{-15}$ erg s$^{-1}$ cm$^{-2}$ flux range (see Tozzi et
al. 2001). Note that this value is meant to include the average effect
of the intrinsic absorption among the AGN population, and it is
different from the adopted value for the slope of the intrinsic
emission $\Gamma = 1.8$. Luminosities are calculated adopting the
photometric redshift of Ilbert et al. (2009) and Salvato et
al. (2011), and the spectroscopic redshift from the zCOSMOS e
Magellan-COSMOS catalogs when available.

Among the 19 FRI candidates observed in the C-COSMOS field, 13 have no
X--ray counterparts in the COSMOS X--ray point source catalog. Among
them, FRI 1 and 4 are not included in the C-COSMOS catalog, probably
due to their proximity with bright sources; despite this, they have
robust X-ray counterparts with about 15 net counts from our aperture
photometry. Uncertainties in the photometric redshifts have been
propagated to the luminosities using a Monte Carlo simulation.  On
average, we have a relative error on the X-ray luminosity of about
30\%, mainly due to poor S/N.

\begin{figure}
\begin{center}
 \includegraphics[width=0.70\columnwidth]{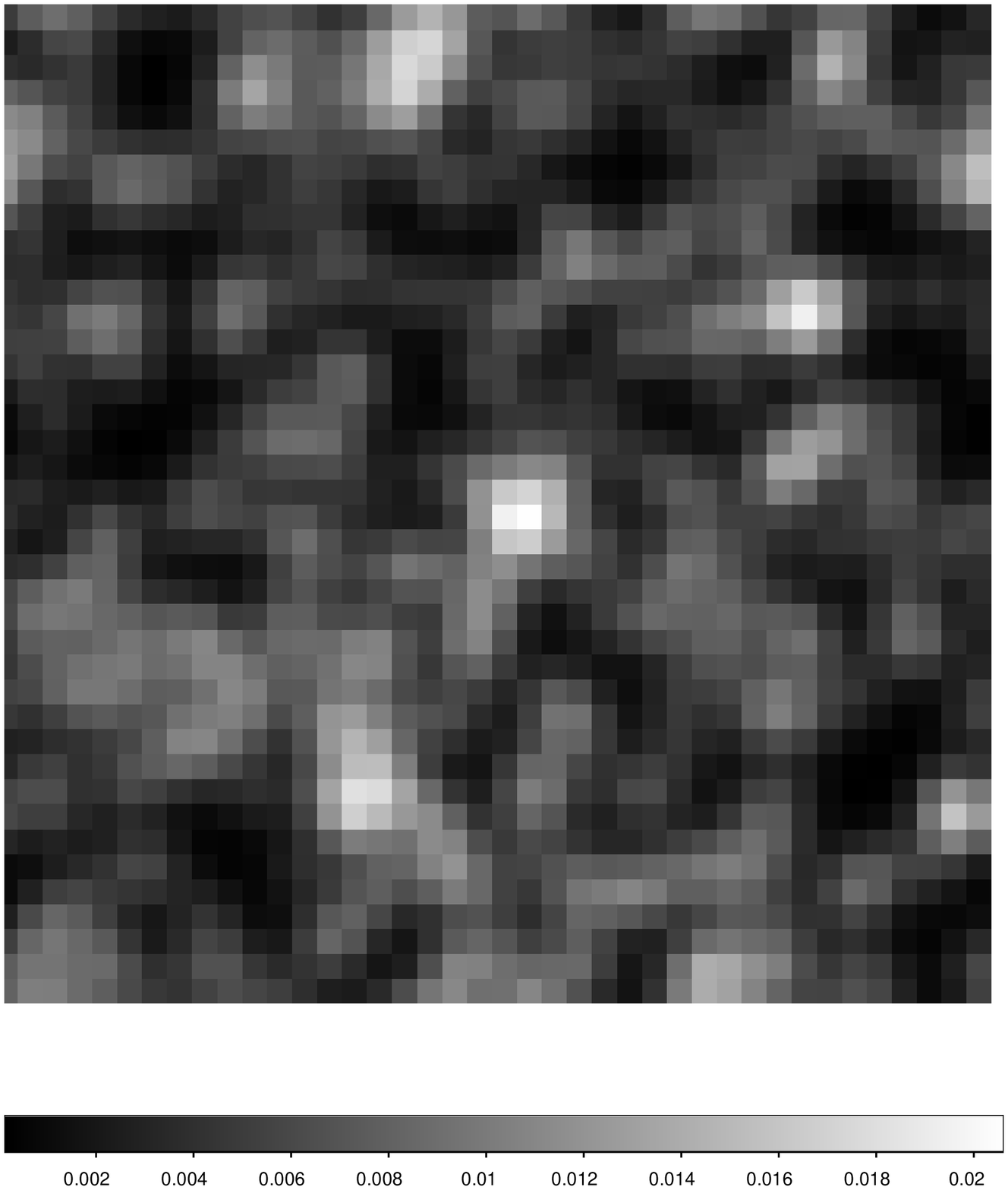}
 \includegraphics[width=0.70\columnwidth]{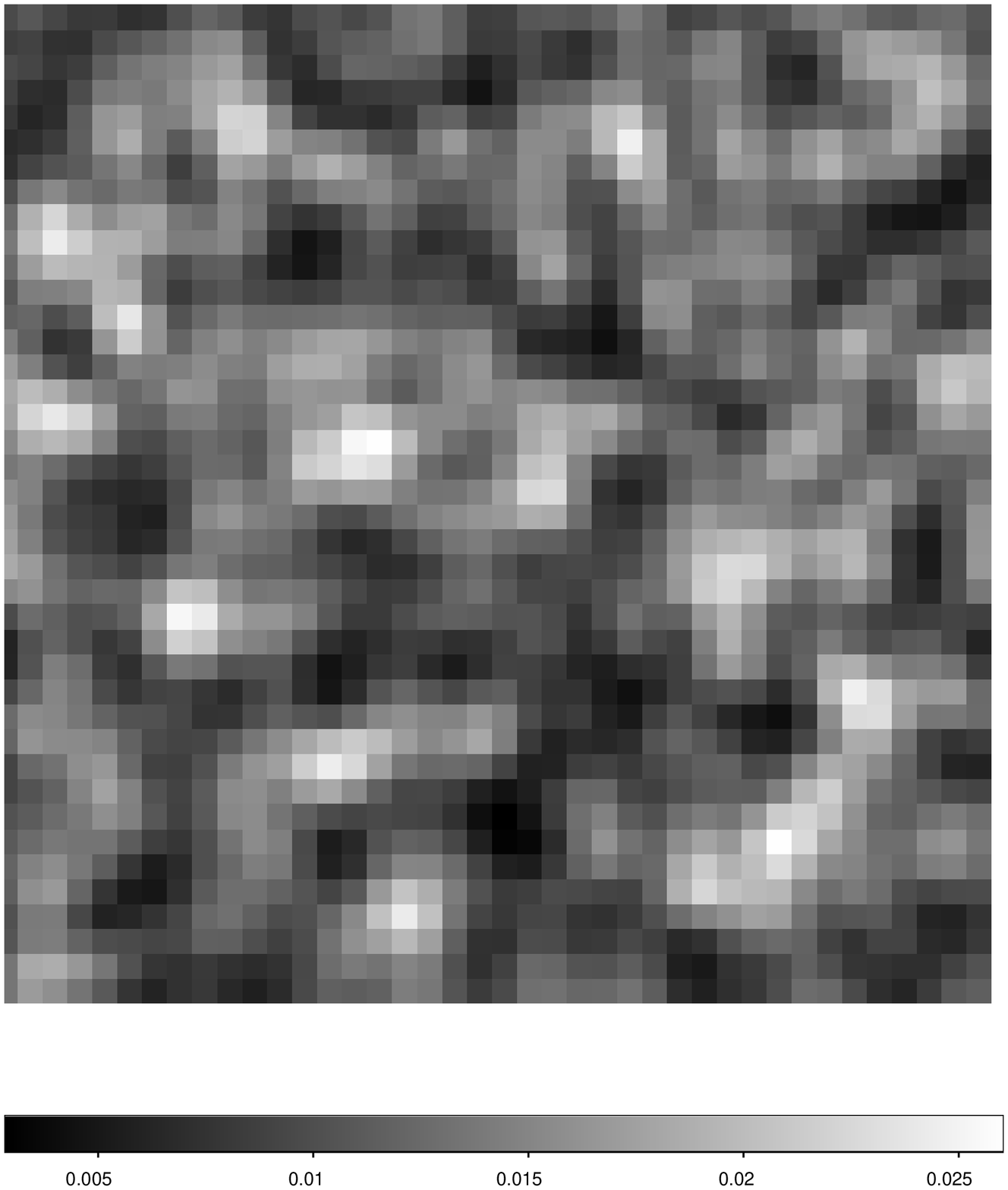}
  \caption{Upper panel: smoothed stacked image in the soft (0.5-2 keV)
    band of the 11 FRI candidates without X--ray detection (see text
    for details).  Lower panel: the same as in the upper panel for the
    2-7 keV band. The size of the images is 40 $\times$ 40 arcsec.}
 \label{st_s_1-4}
\end{center}
\end{figure}

We investigate the X--ray properties of the 11 FRI candidates with no
clearly detected X--ray emission.  We produced soft and hard stacked
X--ray images, shown in Figure \ref{st_s_1-4}, adding the position of
the 11 radio sources after masking all the C-COSMOS X--ray
sources. The stacked images show a signal in the soft band, while none
is visible in the hard band. We tested the robustness of the procedure
by stacking random positions in the masked C-COSMOS image.  We measure
positive emission at a 2 $\sigma$ level in the soft band, with $9 \pm
4$ net counts, and formally zero flux in the 2-7 keV band, ($6 \pm 5$
net counts), in agreement with the single source aperture
photometry. This very weak signal can not be used to infer information
on the spectral shape, nevertheless, it is compatible with no
absorption at all. Since there are 8 FRI candidates with detected
X--ray emission, one would expect to measure a larger X--ray signal in
the stacked images if the average X-ray properties were the same for
all the FRI sample. This difference can not be explained by lower
effective exposure times for the non-detections.  In fact, after
normalizing their count rate to the average exposure time, the stacked
signal of the 11 FRI without X-ray counterpart is only 13 and 9 net
counts in the soft and hard band, respectively.  This shows that the
X--ray properties vary significantly from source to source among the
FRI candidates, with more than half of them showing no X--ray emission
at all.

\begin{table*}
\centering
 \begin{minipage}{140mm}
\caption{Spectral analysis for the FRI candidates with an X--ray
  detected emission with $\simgt$30 Net Counts: (1) FRI Id; (2) C-COSMOS
  Id; (3) total net counts; (4) intrinsic absorption; (5),(6) soft and
  hard fluxes; (7)(8) soft and hard intrinsic luminosity; (9) Goodness
  parameter}
\label{spectroscopy_X_1}
\begin{center}
\begin{tabular}{|ccccccccc|}\hline
 \bf FRI & \bf C-COSMOS  & \bf Net Cts &\bf N$_H$                 &
 \bf S$_{0.5-2}$ &\bf S$_{2-10}$ &\bf L$_{0.5-2}$ &\bf L$_{2-10}$ & \bf goodness \\
 \bf     & \bf           & \bf         & \bf 10$^{22}$ cm$^{-2}$  &   \multicolumn{2}{c}{\bf 10$^{-16}$ erg cm$^{-2}$ s$^{-1}$} & \multicolumn{2}{c|} {\bf 10$^{42}$ erg s$^{-1}$ } &  \\  \hline 
\bf 03  &  451 & 124$^{+25}_{-23}$  &  35$^{+5}_{-5}$    & 11.0$\pm$2.0  & 138$\pm$14 & 196$\pm$35  & 445$\pm$49 & 5\% \\
\bf 05  & 1091 &  34$^{+16}_{-14}$  &   7$^{+6}_{-4}$    &  4.7$\pm$1.3  &  18$\pm$5  &  20$\pm$6   &  42$\pm$12 & 47\% \\ 
\bf 13  & 1242 &  26$^{+16}_{-13}$  &   8$^{+3}_{-3}$    &  3.2$\pm$1.0  &  25$\pm$8  &  10$\pm$3   &  17$\pm$6  & 48\% \\ 
\bf 52  &  701 &  38$^{+17}_{-14}$  & 1.7$^{+0.6}_{-1}$  &  7.3$\pm$1.3  &  24$\pm$7  & 4.3$\pm$1.0 &   6$\pm$2  & 3\% \\ 
\hline
\end{tabular}
\end{center}
\end{minipage}
\end{table*}

We can push further the investigation of the X--ray properties of the
FRI candidates with X--ray detection, with a detailed spectral
analysis using {\tt Xspec} v12.6.0.  Since each source is observed in
several exposures, each spectrum is obtained by adding the spectra
extracted from each observation using the standard FTOOLS (Blackburn
1995) routines. Spectra are extracted from a circular region with a
fixed radius of 5 arcsec. Local background regions were extracted from
annuli around the X--ray sources with an inner and external radii of 7
and 12 arcsec, respectively, after masking nearby X--ray sources to
avoid contamination.  The effective area and response files are
created for each observation and then combined using the FTOOLS
routines ADDARF and ADDRMF.

\begin{figure*}
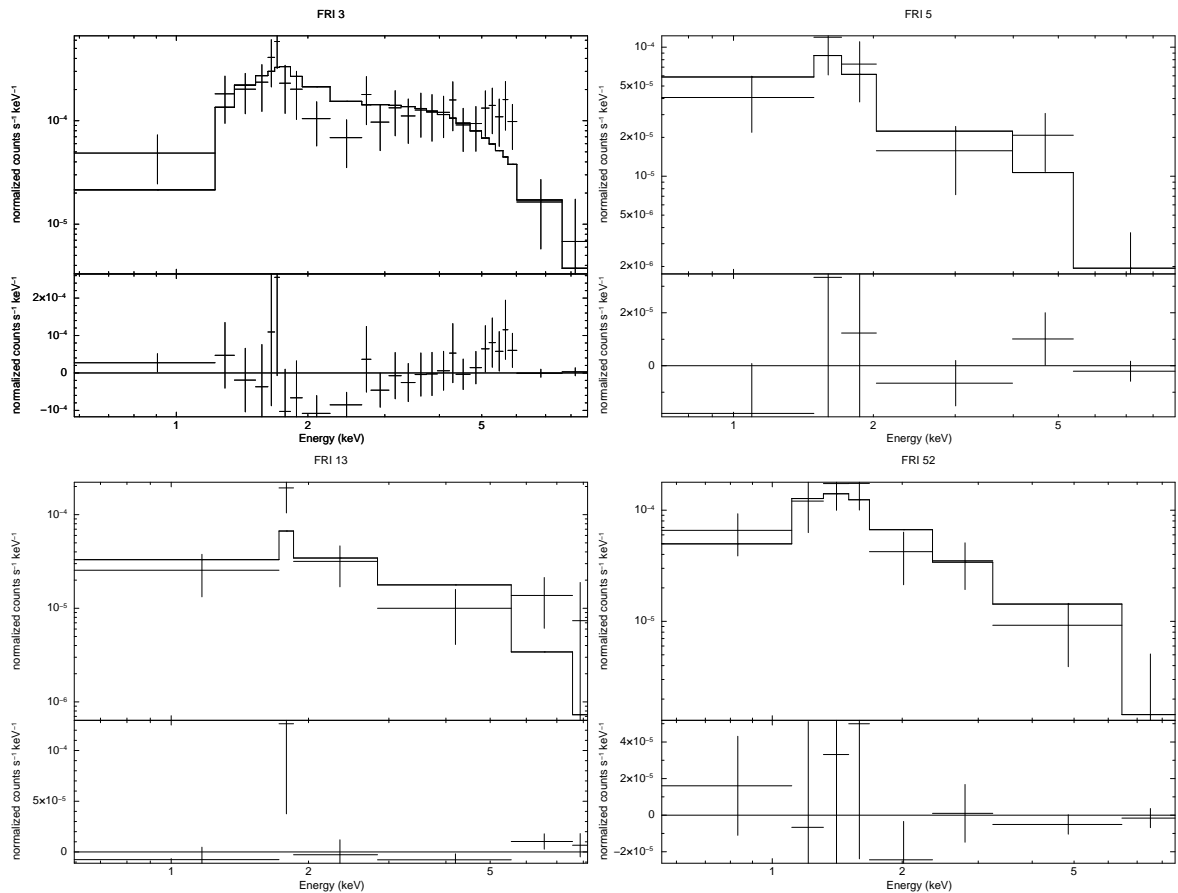

\begin{center}
 \includegraphics[width=0.70\columnwidth, angle=270]{fig3.eps}
 \includegraphics[width=0.70\columnwidth, angle=270]{fig4.eps}
 \includegraphics[width=0.70\columnwidth, angle=270]{fig5.eps}
 \includegraphics[width=0.70\columnwidth, angle=270]{fig6.eps}
 \caption{Spectral data and best--fit models with residuals for FRI
   sources 3 (upper left), 5 (upper right), 13 (bottom left), 52
   (bottom right).}
 \label{fit_X}
\end{center}
\end{figure*}

The adopted spectral model is an absorbed power law, with a fixed
Galactic absorption. The value of the Galactic absorption is $N_H= 1.7
\times 10^{20}$ cm$^{-2}$, obtained as the average value in the
C--COSMOS region measured in the Leiden/Argentine/Bonn survey
(Kalberla et al. 2005).  Given the low S/N of our spectra, the
intrinsic power law index is fixed to $\Gamma=1.8$, which is the
average value found in AGN in a similar flux range (see, e.g., Tozzi
et al. 2006). FRI 1, 4, 18 and 66 are detected with a number of net
counts too low (about 10-15 in the 0.5-7 keV band) to have a
meaningful spectral analysis. We note also that these sources have the
lowest exposure time among the sources with an X--ray
counterpart. Best fit parameters are found by minimizing the Cash's
$C$--statistic (Cash, 1979; Nousek \& Shue, 1989) which is appropriate
for low S/N. We also report a {\sl goodness} parameter to evaluate the
quality of the fit performed within {\tt Xspec} by means of Monte
Carlo
simulations\footnote{http://heasarc.gsfc.nasa.gov/docs/xanadu/xspec/manual/XSgoodness.html}.
When the best-fit is acceptable, the {\sl goodness} parameter is
around 50\%.  In Table \ref{spectroscopy_X_1} we report the results of
our spectral analysis.  In Figure \ref{fit_X} we show the spectral
fits for FRI candidates 3, 5, 13 and 52.

Fluxes derived from fits to the spectral template are in agreement
with the fluxes obtained directly from the count rate with an average
conversion factor (see Table \ref{photometry_X}).  All the examined
sources are consistent with being absorbed at the level of a few
$\times$10$^{22}$ cm$^{-2}$, with source FRI 3 having $N_H=$
3.5$\pm$0.5$\times$10$^{23}$ cm$^{-2}$.

We further explore the spectrum of FRI 3, which is the source with the
largest number of counts. We relax the constraints on the intrinsic
power law $\Gamma$, finding $\Gamma = 1.6 \pm 0.8$ and
$N_H=$2.6$\pm$0.4$\times $10$^{23}$ cm$^{-2}$, with a slight
improvement in the quality of the fit.  The best fit obtained with a
reflection-dominated spectral model ({\tt PEXRAV}) shows only a modest
increase in the quality of the fit, but with an unrealistic value for
the intrinsic slope $\Gamma \sim 0$.

Given the poor S/N of sources FRI 5, 13 and 52, the errors on the
measured $N_H$ are quite large, and all the best-fit values are
consistent with no absorption within 2 $\sigma$.  In order to get a
lower limit on the average absorption, we stack the signal of all the
radio sources with X-ray counterpart with the exception of FRI 3.  We
obtain a total of 88$\pm$10 and 73$\pm$11 net counts in the soft and
hard bands, respectively. This correspond to $HR=-0.09\pm 0.09$.
Since the average redshift of this 7 sources is
$\langle$z$\rangle=$1.09, we can estimate the average absorption by
comparing the $HR$ values seen by Chandra as a function of intrinsic
$N_H$ for that redshift.  We find an average $N_H=$3.1$\pm$1.2 $\times
10^{22}$ cm$^{-2}$.  This shows that the average intrinsic absorption
the in the X--ray detected FRI sources is $N_H \geq 10^{22}$
cm$^{-2}$) at 2 $\sigma$ confidence level.

Therefore, we find that at least some of the FRI candidates, namely
those with X--ray counterparts in the C-COSMOS data, show sign of
significant intrinsic absorption, at a level typically higher than
those found in local FRIs (Donato et al. 2004, Balmaverde et
al. 2008).  Donato et al. (2004) found 8 FRI sources out of 25 with
excess absorption over the Galactic value with rest-frame column
densities N$_H\sim$10$^{20}$--10$^{21}$ cm$^{-2}$, and no sources with
N$_H\sim$10$^{22}$ cm$^{-2}$. The power law photon index of their
sample spans in the range of values $\Gamma \sim$1.1--2.6, with
average value $\Gamma$=1.9 and standard deviation $\sigma$= 0.4.
Balmaverde et al. (2008) detected nuclear X--ray absorption at a level
of N$_H\sim (0.2-6)\times 10^{22}$ cm$^{-2}$ in 4 out of 18 FRI
sources.  They observe that X--ray absorbed sources in their sample of
low-luminosity 3C/FRI galaxies are associated with the presence of
highly inclined dusty discs seen in the HST images, and suggest the
existence of a flattened X--ray absorber, but of much lower optical
depth than in classical obscuring tori. In their sample, they found a
range of $\Gamma \sim$0.8--2.4, again with average value $\Gamma$=1.9
and standard deviation $\sigma$= 0.5.

X--ray unabsorbed luminosities span a range of values larger than
those observed in low redshift FRIs. Donato et al. (2004) report just
7 out of 25 sources with luminosities $L_X\leq 10^{42}$ erg sec$^{-1}$
in the 0.3--8 keV band, and only 2 with luminosities $L_X\sim 10^{43}$
erg sec$^{-1}$.  Balmaverde et al. (2008) report only one FRI source
with $L_X\simgt 10^{43}$ erg sec$^{-1}$ in the 0.5--5 keV band.

\begin{table}
\centering
\begin{minipage}{85 mm}
\caption{Radio Power for the FRI candidates: (1) FRI Id; (2) C-COSMOS
  Id (if present); (3) observed FIRST flux (mJy) at 1.4GHz, from Table
  1 in Chiaberge et al. (2009); (4) rest frame luminosity at 1.4GHz.}
\label{radio}
\begin{tabular}{cccc}\hline
 \bf FRI & \bf C-COSMOS  & \bf S$_{1.4GHz}$ & \bf  L$_{1.4GHz}$  \\
 \bf     & \bf           & \bf       mJy           & \bf  10$^{32}$ erg s${^-1}$ Hz$^{-1}$\\  \hline
\bf 07  &      &  5.99  &     6.706 \\            
\bf 11  &      &   1.3  &     2.543 \\            
\bf 16  &      &  1.14  &    0.4483 \\            
\bf 20  &      &  1.13  &     0.985 \\            
\bf 22  &      &  1.51  &     0.853 \\            
\bf 26  &      &   5.7  &     2.417 \\            
\bf 28  &      &  4.39  &     1.627 \\            
\bf 29  &      &  1.33  &    0.6055 \\            
\bf 30  &      &  2.74  &     5.063 \\            
\bf 31  &      &  1.88  &     1.329 \\            
\bf 32  &      &  1.77  &     7.725 \\  \hline
\bf 01  &      &  1.79  &    0.6074 \\            
\bf 03  &  451 &  2.12  &     2.952 \\            
\bf 04  &      &  4.21  &     9.528 \\            
\bf 05  & 1091 &  1.26  &    0.4523 \\            
\bf 13  & 1242 &  3.71  &     1.364 \\            
\bf 18  &  427 &  1.31  &    0.1628 \\            
\bf 52  &  701 &  1.54  &    0.3443 \\            
\bf 66  &  669 &  1.11  &    0.2043 \\   \hline   
\end{tabular}
\end{minipage}
\end{table}

To summarize, the FRI candidates presented in Chiaberge et al. (2009)
show X--ray characteristics that vary significantly from source to
source. In this sample 11 sources out of 19 have no X--ray
counterpart, but they are detected at a 2 $\sigma$ confidence level in
their stacked soft-band image, while they are still not detected in
their stacked hard band image.  Therefore, they do not show any hint
of being absorbed AGN. On the other hand, radio sources with X--ray
counterpart show significant intrinsic absorption from single-source
spectral analysis, and the hardness ratio of the stacked image
suggests an average $N_H \simeq 10^{22}$ cm$^{-2}$.  Among them, FRI
sources 3, 5 and 13 show higher intrinsic luminosities, while FRI 1,
4, 18, 55 and 66 have luminosities comparable to that of low-z FRIs.
Both luminosities and intrinsic absorption reach values that are
unfrequently observed in low-z sources.

However, we note that we are studying a sample that might include a
few objects with radio power more typical of the FRII population. The
1.4GHz radio luminosity of the FRI candidates in our sample is shown
in Table \ref{radio}.  We assumed a power law energy index $\alpha
\sim 0.8$.  We note that among the sources with X--ray counterpart,
two are within a factor of 2 of the FRI$/$FRII break luminosity
$L_{break}(1.4GHz) \sim 4 \times 10^{32}$ erg~s$^{-1}$~Hz$^{-1}$,
while the remaining 4 have radio luminosities ten times lower than
$L_{break}$.  The highest absorption and luminosity is measured for
the radio source with $L_{1.4GHz} \sim L_{break}$.  Despite our sample
may include some FRIIs, we cannot exclude a possible evolution of the
X--ray properties of FRI sources, suggesting that the accretion in
high redshift FRIs may be different with respect to low redshift ones.
In particular, the presence of a dusty torus in high redshift FRIs
could explain the fraction of highly absorbed sources.  In low-z FRIs,
the presence of a dusty torus is usually rejected on the basis of the
low intrinsic absorption typically measured at X--ray wavelengths
(Donato et al. 2004), and of the high optical-core detection rate
(Chiaberge, Capetti \& Celotti 1999). However, several works
(Hardcastle et al. 2002; Evans et al. 2006) point out that a jet
dominated X--ray emission occurring on scales larger than a torus
(therefore mimicking low absorption) cannot be excluded, and this
leaves room to the presence of the torus.  Another possible
explanation is that we may be observing a sort of intermediate
population, with characteristics that are in-between FRIs and FRIIs,
with a radio morphology typical of low redshift FRIs and with X--ray
properties more typical of FRIIs.  Such sort of intermediate objects
are already known (Capetti, Fanti \& Parma 1995).

\section{Search for X--ray extended emission}

Under the assumption that the environment of FRIs does not change with
redshift, high-z FRIs can be used as signposts for high-z clusters.
This aspect can be particularly relevant for redshift z$\sim$1.5 and
higher, an epoch still out of reach of current cluster surveys, both
in the IR and in the X--ray.  The typical flux level of the ICM from a
hot (kT > 5 keV) clusters at z$\sim$1.5 is $F_{0.5-2 keV} \sim 1-2
\times 10^{-14}$ erg s$^{-1}$ cm$^{-2}$; the stacking technique
virtually multiply the effective exposure time by the number of FRI
sources, so that the stacked ICM emission, if present, would reach
fluxes as high as $10^{-13}$ erg s$^{-1}$ cm$^{-2}$.

\begin{figure}
\begin{center}
 \includegraphics[width=0.70\columnwidth]{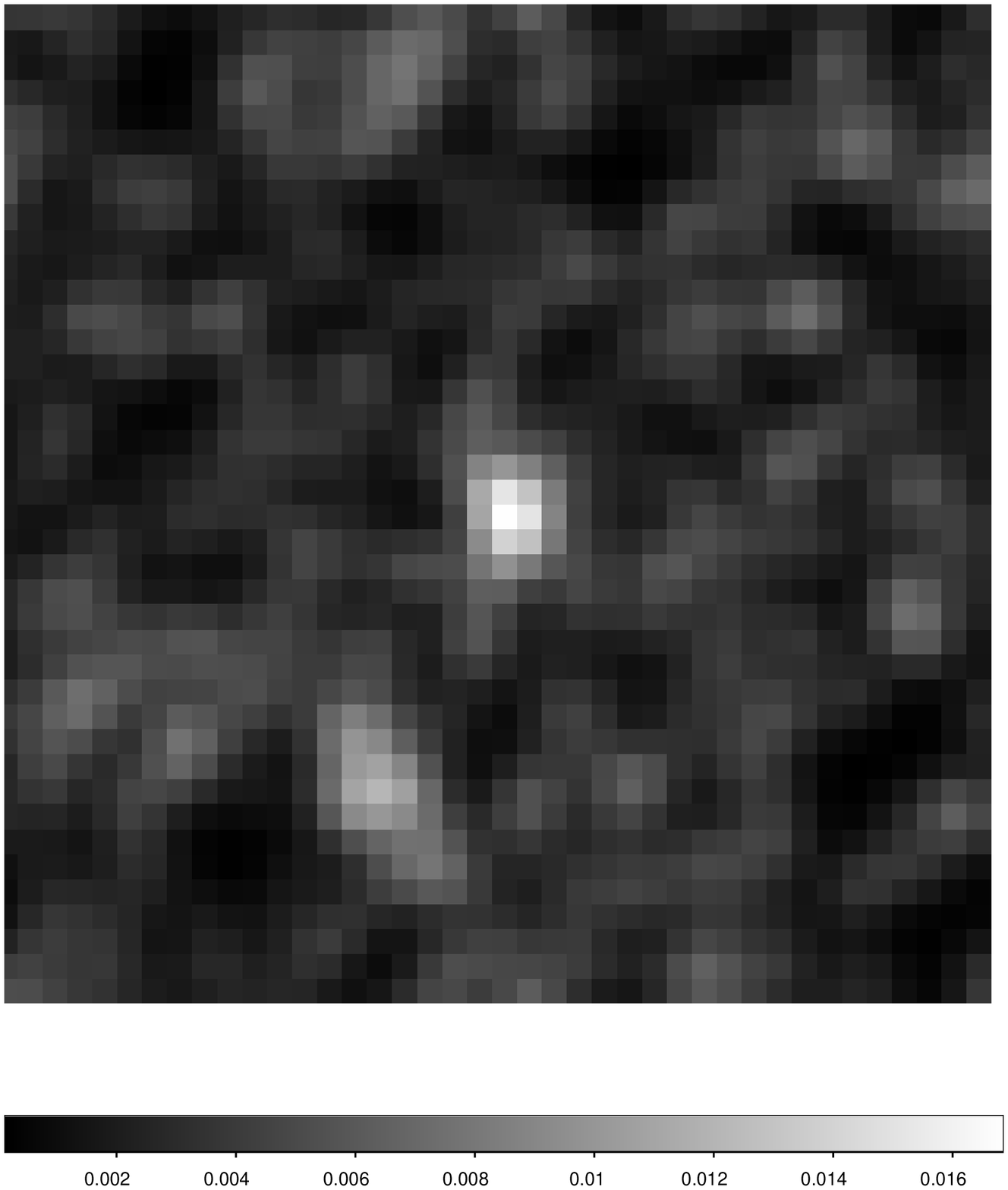}
 \includegraphics[width=0.70\columnwidth]{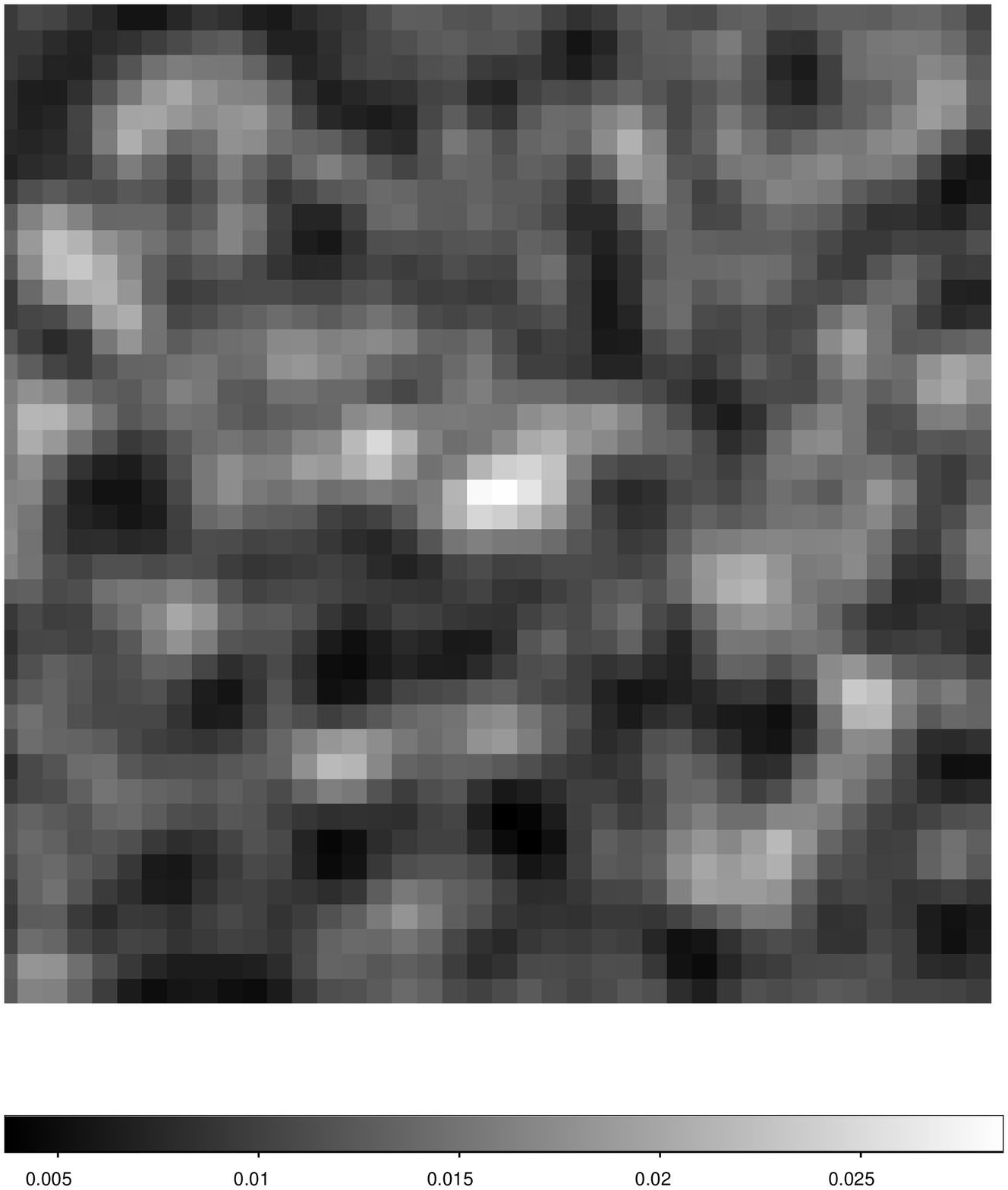}
 \caption{Smoothed stacked images in the soft (upper panel) and hard
   band (bottom panel) of all the 19 FRI candidates. The size of
     the images is 40 $\times$ 40 arcsec. }
 \label{st_all}
\end{center}
\end{figure}

\begin{figure}
 \includegraphics[width=\columnwidth]{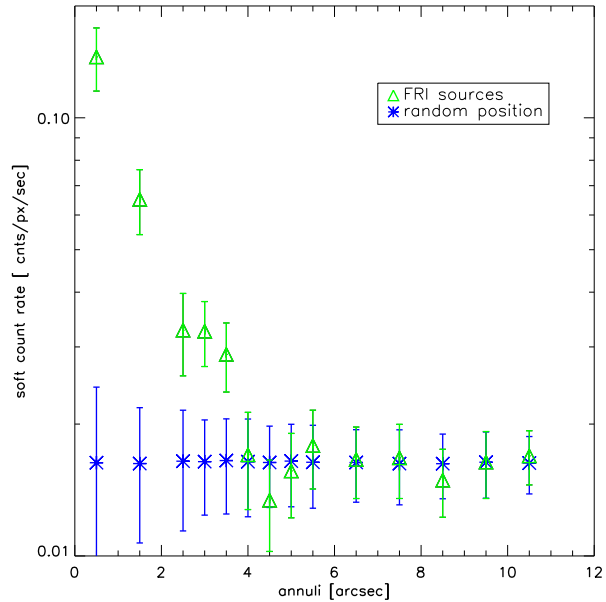}
 \caption{Radial profile in the soft band of the staked image of the
   19 FRI sources (green triangles) compared with the radial profile
   measured in random fields (blue asterisks).}
 \label{profile}
\end{figure}

We create stacked images using all the 19 FRI candidates; the stacked
images are shown in Figure \ref{st_all}.  A visual inspection shows no
clear signs of extended emission in the soft band.  The same is found
in the hard band, where the ICM emission is expected to be much weaker
in any case.  To provide a quantitative measure, we compute the count
rate in the soft band in annuli with inner and outer radii separated
by 1 arcsecond. Then we compare these values to those found in images
obtained stacking random positions from the masked image.  Figure
\ref{profile} show the soft-band count rate in each annulus of the
stacked image of the FRI sources versus the mean value measured in the
control-image.  We find evidence for emission only within 4 arcsec,
while the emission at larger radii is consistent with noise.  The
stacked image is consistent with unresolved sources imaged at
different off axis angle, due to an average Point Spread
Function with FWHM of $\sim 2.5$ arcsec.  

\begin{figure*}
\begin{center}
 \includegraphics[width=0.70\columnwidth]{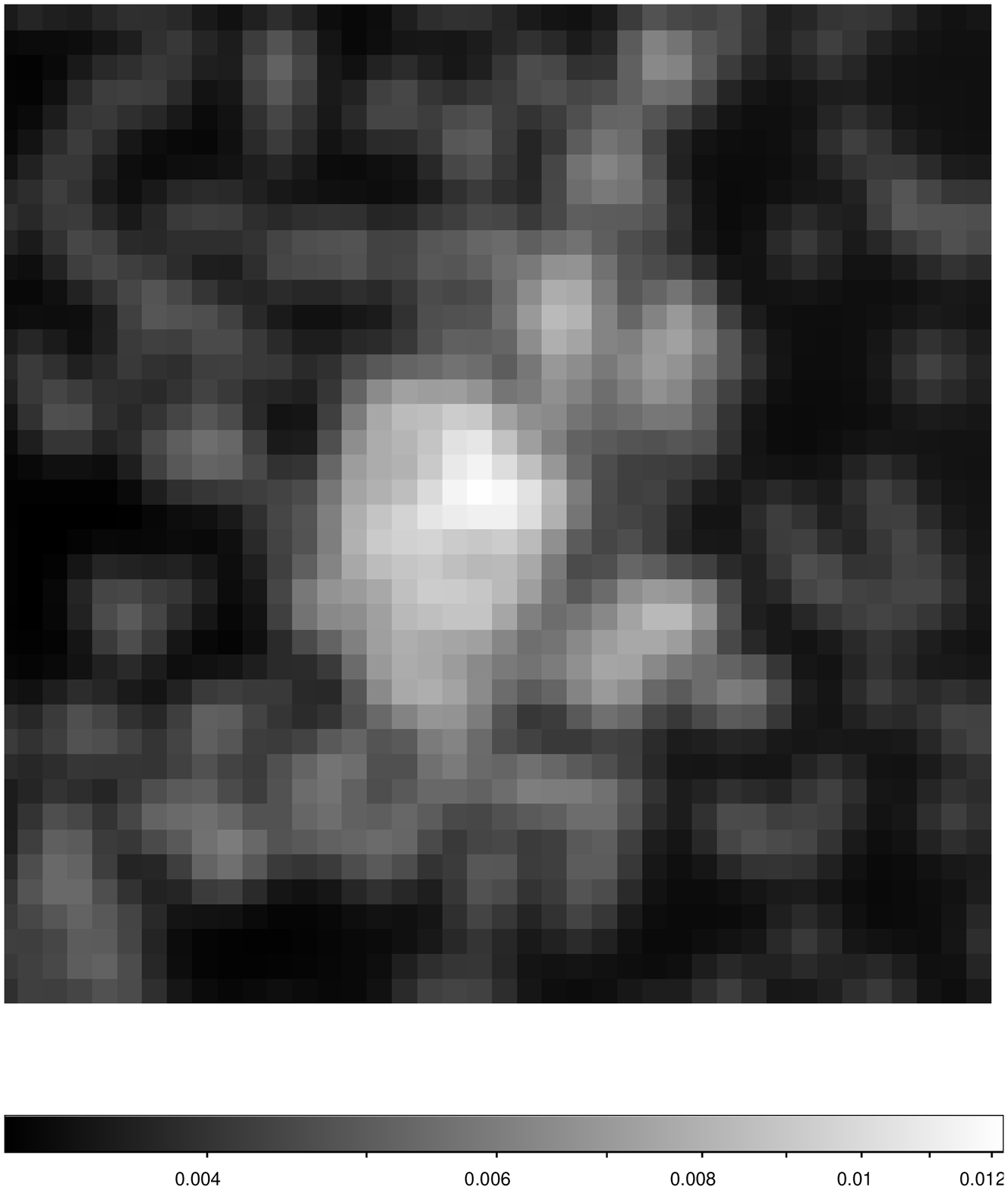}
 \includegraphics[width=0.85\columnwidth]{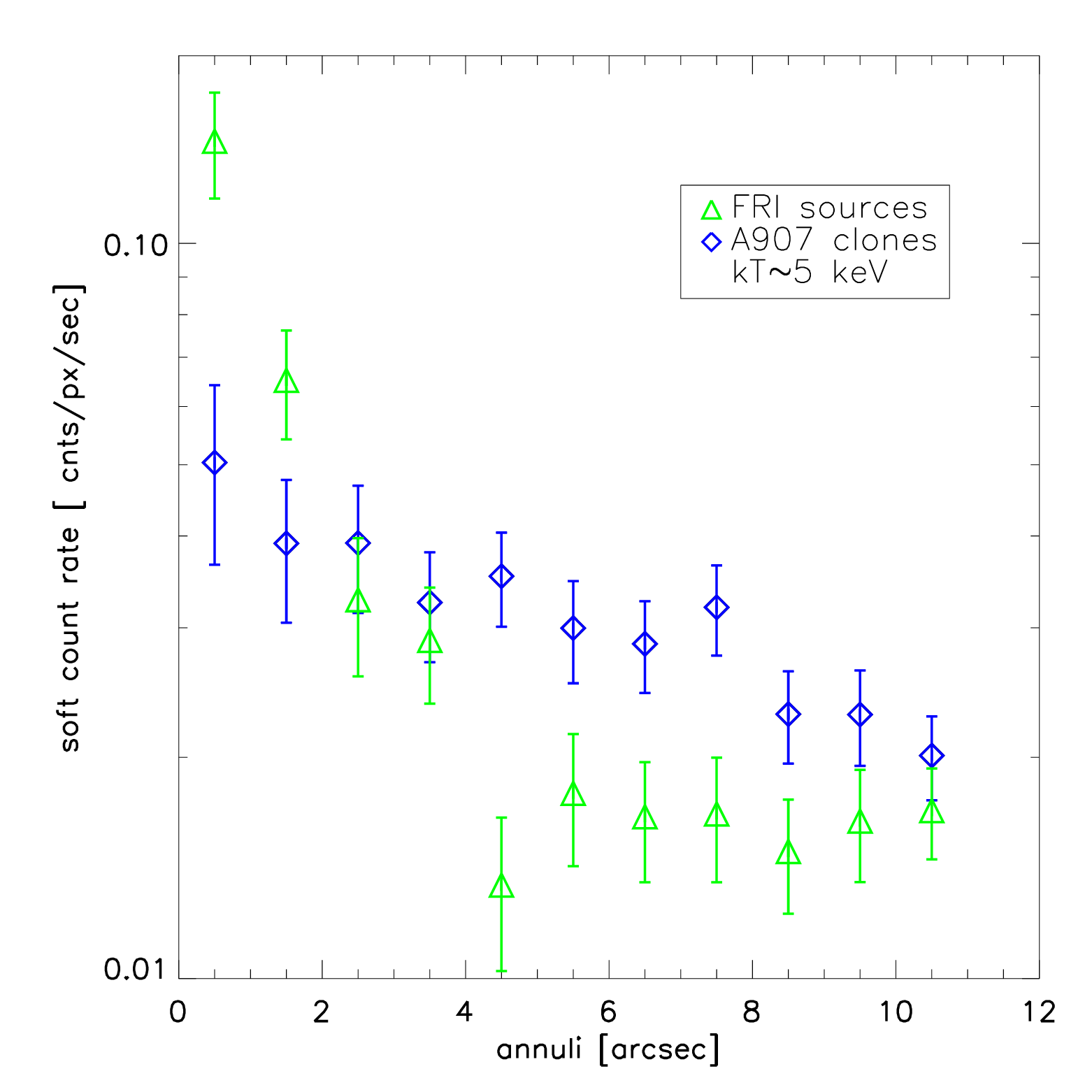}
 \includegraphics[width=0.70\columnwidth]{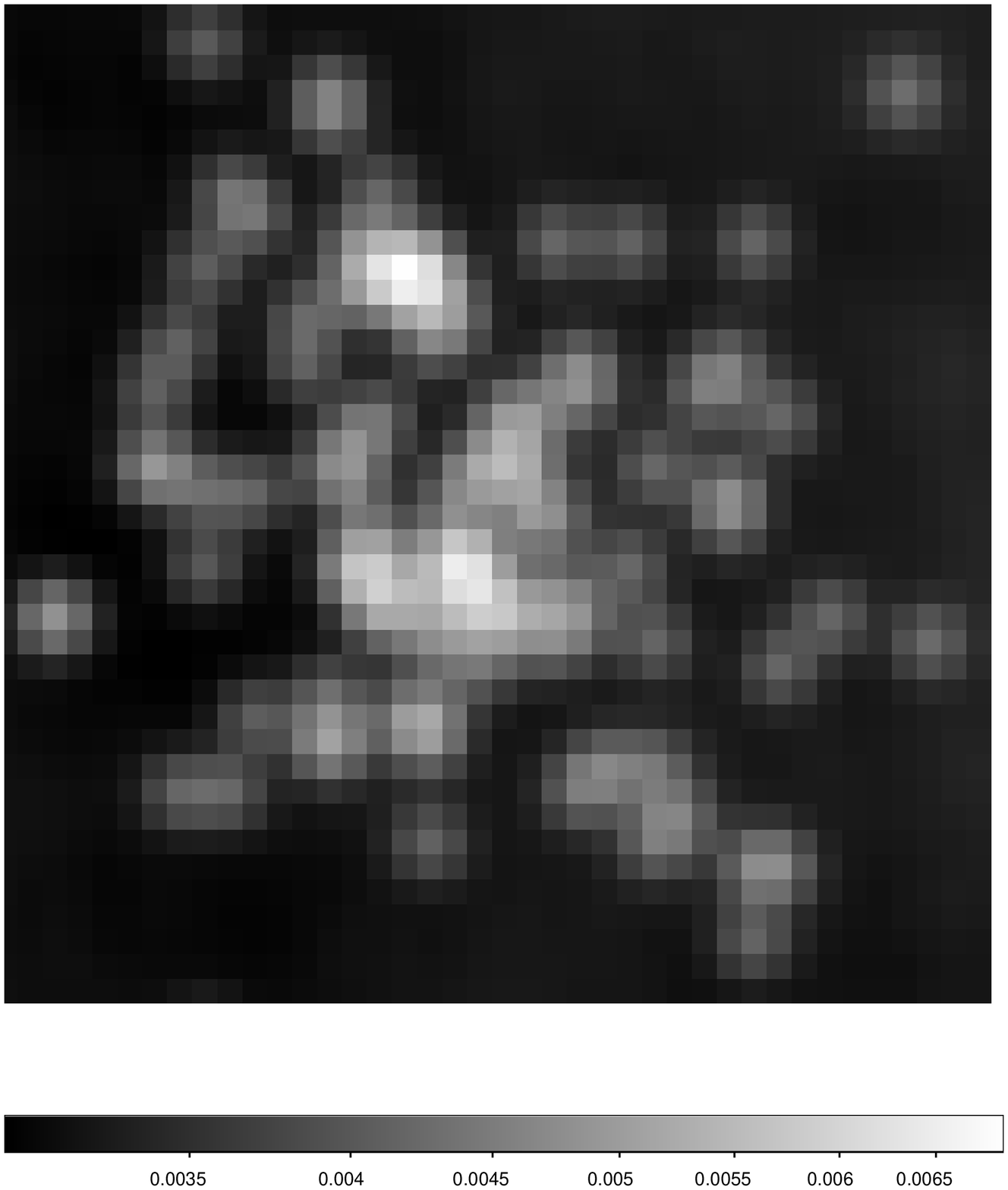}
 \includegraphics[width=0.85\columnwidth]{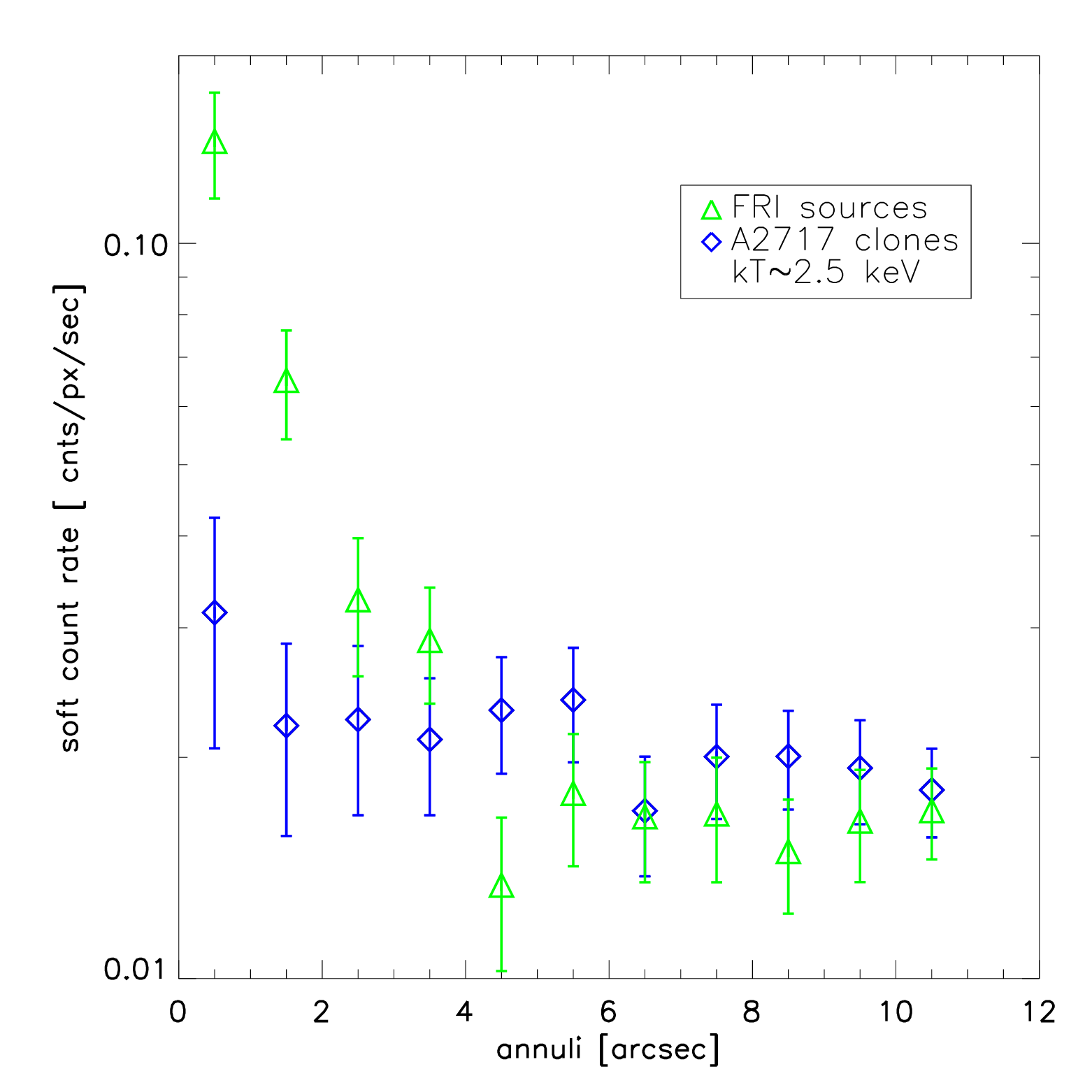}
 \includegraphics[width=0.70\columnwidth]{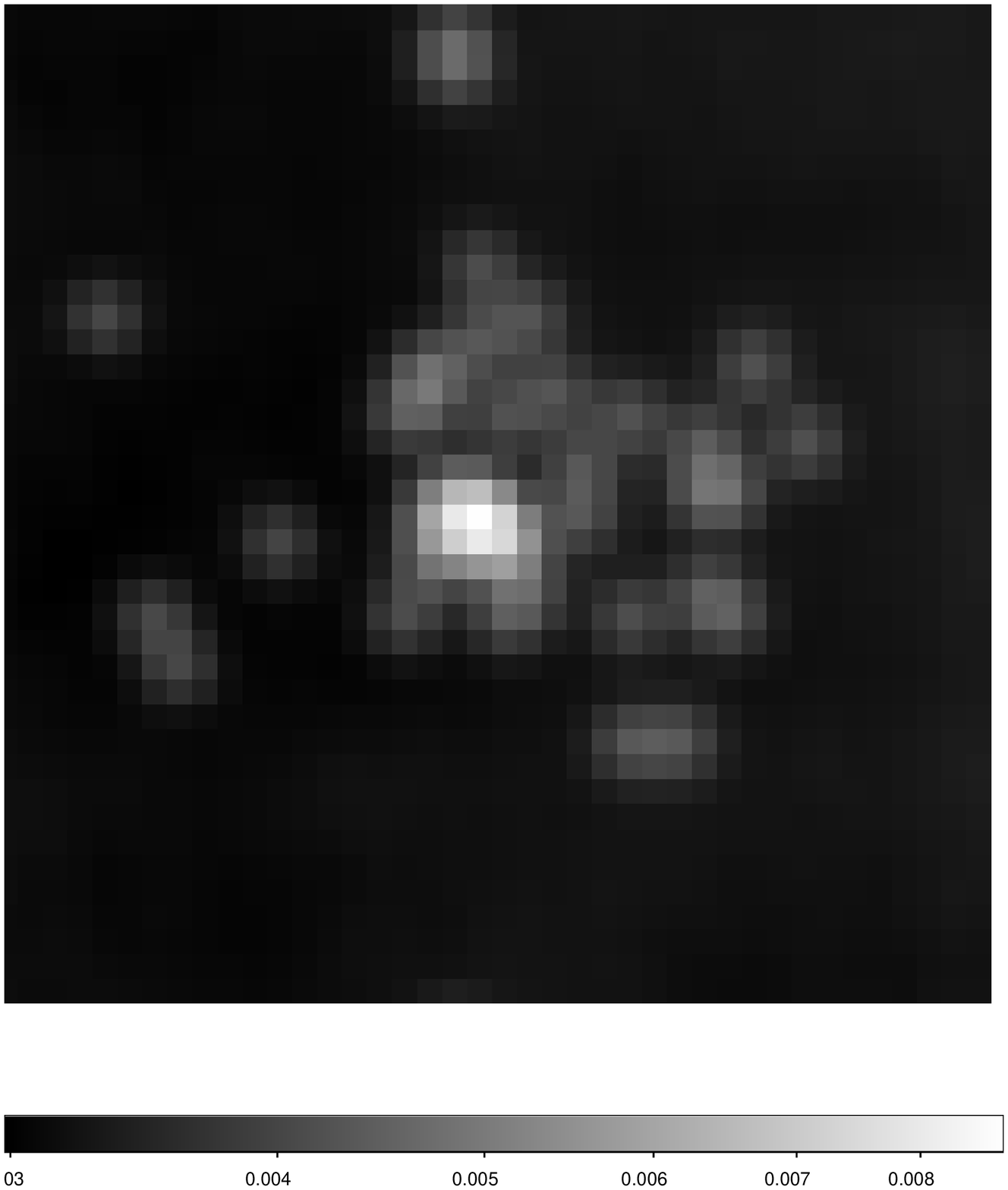}
 \includegraphics[width=0.85\columnwidth]{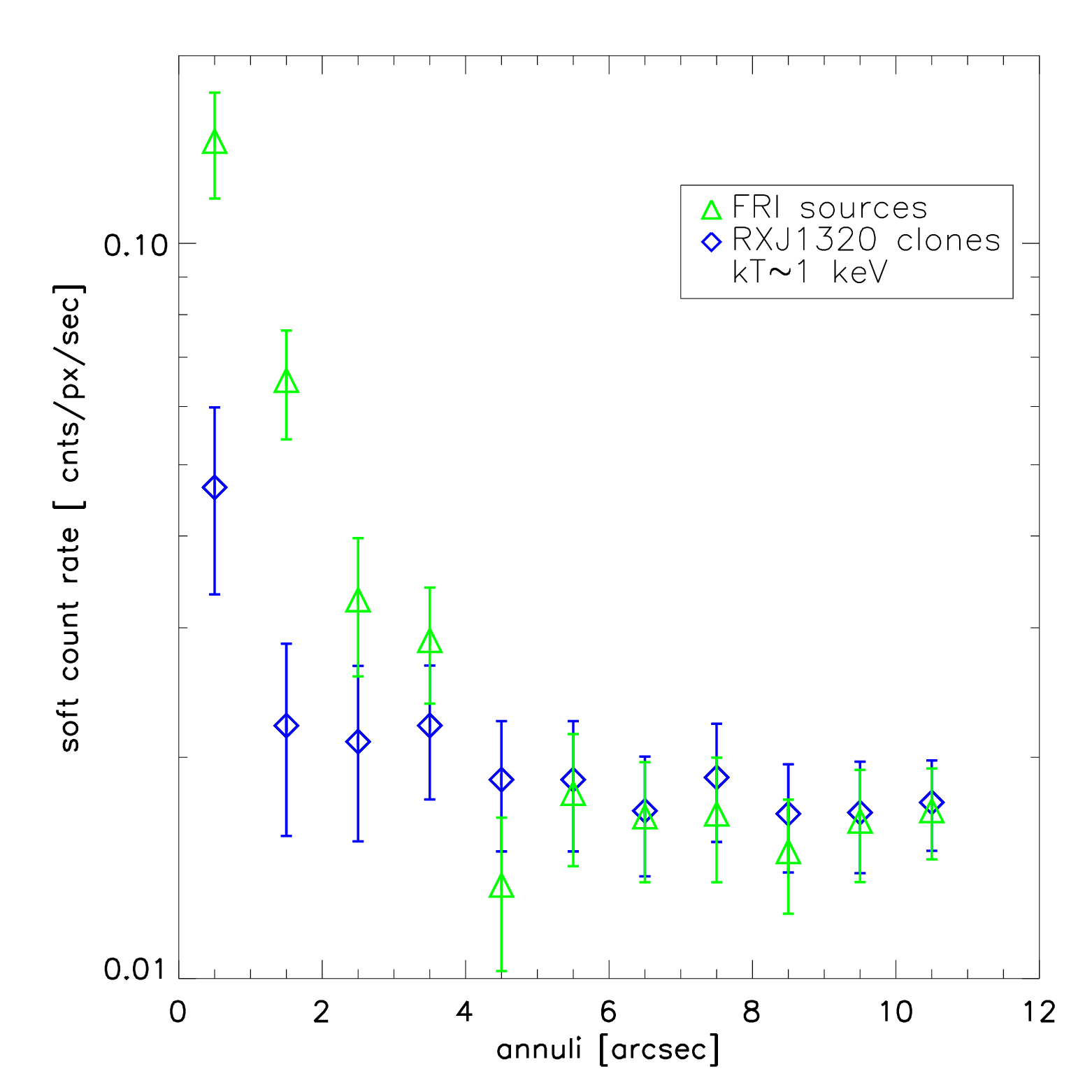}
\caption{Upper left panel: soft-band, smoothed stacked image assuming
  virialized haloes with $kT\simeq 5$ keV in the position of the FRI
  sources; upper right panel: count rate radial profile for the mock
  image (blue diamonds) compared with the radial profile found in the
  stacked image of the FRI sources (green triangles).  Middle panels:
  the same as the upper panels assuming virialized haloes with
  $kT\simeq 2.5$ keV.  Bottom panels: the same as the upper panels
  assuming virialized haloes with $kT\simeq 1$ keV. The size of the
  images is 40 $\times$ 40 arcsec. }
 \label{clones}
\end{center}
\end{figure*}

However, the failed detection of extended emission does not rule out
the presence of a virialized halo around the FRI sources.  To put a
robust upper limit to the extended emission that can be allocated
around the FRI candidates, we proceed as follows. We associate to each
FRI source the typical X--ray emission of a group or cluster at a given
temperature.  In order to model this ICM emission, we use the Chandra
images of real clusters and groups of galaxies cloned to the redshift
of the FRI sources.  This technique has been already used to
investigate the evolution of cool cores at high redshift (see Santos
et al. 2008).  We used 3 different clusters as template: A907 with a
temperature of 5 keV and a luminosity of $4.7\times 10^{44}$ erg
s$^{-1}$ (typical of rich clusters); A2717 with a temperature of 2.4
keV and a luminosity of $0.78 \times 10 ^{44}$ erg s$^{-1}$ (typical
of small clusters); and RXJ1320 with a temperature of $\sim 1$ keV and
a luminosity of $0.26\times 10^{44}$ erg s$^{-1}$ (appropriate for
galaxy groups).  The cluster image is cloned to the corresponding FRI
redshift and rotated randomly.  The nuclear emission is not added to
the mock image.  The measured count rate for the stacked clones in the
annuli, compared to the result from FRI stacked image, is shown in
Figure \ref{clones}.  The top, middle and bottom panels show the
images obtained using clusters A907, A2717 and RXJ1320, respectively.

In the hypothesis that FRIs are inside clusters with typical
temperature around 5 keV (top panels), the stacked image of 19 of
those would have shown significant emission at large radii. Therefore,
we can rule out such an hypothesis.  The mock image obtained with
clones of low temperature clusters ($kT\sim 2.5$ keV) shows some
excess with respect to the FRI stacked image at radii larger than 4
arcsec.  Still, the two profiles do agree within the uncertainties,
also considering that in the local universe about 70\% of the FRI
galaxies are found in clusters (Zirbel 1996). As a consequence, these
results do not rule out the possibility of some of the FRIs being at
the centre of medium cluster of typical temperature $kT \sim 2-3 $
keV. Finally, the mock image corresponding to groups of about $1$ keV
(see bottom panels of Figure \ref{clones}) is consistent with the real
stacked image, since only the innermost bin would bring the signal of
the faint ICM associated to high-redshift groups.  In this last case
the emission observed in the FRI stacked image in the first bins must
be in part associated to the ICM emission of the putative virialized
group surrounding the FRI source.  The fraction of ICM emission is
predicted to be about 20 per cent.  This is still compatible with the
spectral analysis of FRI sources, despite the detection of intrinsic
absorption shows that the majority of the X--ray emission in the inner
5 arcsec must be due to nuclear activity.

To summarize, we obtained constraints to the presence of virialized
haloes with ICM emission around FRIs in this sample.  There is no room
for virialized haloes with $kT\simeq 5$ keV, while group-sized, low
temperature ($kT \leq 2$ keV) ICM emission appears to be consistent
with the X--ray observations of our FRI sample.  Much deeper X--ray
data are needed to search for this faint emission around FRIs.  A
similar project in the CDFS 4Ms field (Xue et al. 2011) is currently
underway.

\section{Conclusions}

We analysed the X--ray properties of a sample of 19 high redshift
(1$<$z$<$2) FRI candidates selected on the basis of the radio flux
(Chiaberge et al. 2009) and observed in the Chandra COSMOS field.
From the spectroscopic and photometric redshift found in the
literature, we find that the rest frame 1.4GHz radio luminosities of
our 19 FRI candidates span a range from $\sim$ 2 $L_{break}(1.4GHz)$
down to $\sim$0.05 $L_{break}$.

We find that 6 of our radio sources have an unresolved X--ray
counterpart in the C-COSMOS catalog.  Two additional radio sources
have a significant X--ray detection in our aperture photometry, even
though they are not included in the C-COSMOS catalog.  We also search
for unresolved X--ray emission below the detection threshold by
stacking the X--ray images in the position of the 11 FRI radio sources
without X--ray counterparts.  We find a 2 $\sigma$ detection in the
soft band, and none in the hard band.

We performed the X-ray spectral analysis of the FRI galaxies with
X-ray counterpart.  We find that source FRI 3, which has the highest
S/N, has a significant intrinsic absorption N$_H = (3.5 \pm 0.5)
\times 10^{23}$ cm$^{-2}$.  The tentative spectral analysis of the
other X-ray sources suggest intrinsic absorption in the range
$10^{22}-10^{23}$ cm$^{-2}$, as suggested also by the HR of their
stacked signal.  However, we can not infer that FRI galaxies at high
redshift preferentially have X-ray absorbed spectra, since such a
conclusion should be based on a complete FRI sample and on deeper
X-ray data.

The X--ray luminosities of the FRI candidates show values that,
although occasionally observed in low redshift FRIs (Donato et
al. 2004, Balmaverde et al. 2008 ), appear to be higher than the
typical FRI X--ray luminosities in the local universe. This finding
should be regarded as tentative since it is based on a sample with 11
X-ray upper limits out of 19 sources. Nevertheless, our study suggests
an evolution of the X--ray properties of FRI sources with redshift,
with accretion at high redshift possibly being more efficient than in
low redshift ones.

We also search for extended X--ray emission possibly associated to the
ICM of the virialized haloes hosting the FRIs.  We compute radial
profiles in several bins around the FRI positions, and find that the
stacked image is consistent with unresolved emission, with no signs of
diffuse X--ray signal.  To put an upper limit to the ICM emission
around FRIs, we compute mock images assuming the presence of a
virialized halo of 5, 2.5 an 1 keV at the position of each FRI.  The
mock images are obtained by stacking the images of real Chandra
clusters, each cloned at the redshift and with an exposure matching
those of the corresponding source.  We exclude the presence of high
temperature virialized haloes with $kT\simeq 5$ keV.  On the other
hand, there is still room for ICM emission from medium and low
temperature ICM ($1-2.5$ keV).

The results of this study show that the X--ray analysis of FRI sources
at high redshift can shed light on several aspects, in particular the
physics of accretion processes in FRIs and the presence of virialized
haloes hosting the FRI galaxies.  However, in order to confirm the
hint of higher intrinsic absorption and higher luminosities in high-z
FRIs, and further explore the presence of ICM around them, we need
deeper X--ray data and larger FRI sample. Considering the stacked
signal of the non X-ray detected sources, an exposure of 1 Msec with
Chandra would guarantee at least the detection ($\sim 10$ net counts)
of the majority of the sample and a robust X-ray spectral analysis
($\geq 100$ net counts) for at least 1/3 of the sample. The best field
where this study can be extended is the Chandra Deep Field South with
4Ms of Chandra time, which has, however, a ten times smaller solid
angle. This study is currently underway.

\section*{Acknowledgments}

We acknowledge support from ASI-INAF I/088/06/0 and ASI-INAF
I/009/10/0.  PT acknowledges support under the grant INFN PD51.  We
thank Joana Santos for providing the high-z clones of local X--ray
clusters observed with Chandra.  We thank Gianluca Castignani for the
providing the radio luminosity of the FRI sample, and Piero Rosati for
helpful discussions.  We also thank Francesca Civano e Giorgio
Lanzuisi for discussion on the spectral analysis of C-COSMOS sources.
Finally, we thank the anonymous referee for detailed and helpful
comments.

\bsp

\label{lastpage}

\end{document}